\documentstyle[12pt,aaspp4]{article}

\lefthead{Bekki \& Couch}
\righthead{Formation  of super star clusters}

\begin{document}
\title{Potential formation sites  of super star clusters in ultra-luminous
infrared galaxies}

\author{Kenji Bekki \&  Warrick J. Couch} 
\affil{School of Physics, University of New South Wales, Sydney 2052, Australia}

\begin{abstract}

Recent observational results on high spatial resolution images 
of ultra-luminous infrared galaxies (ULIGs) have revealed
very luminous, young, compact, and heavily obscured super star clusters 
in their  central regions, suggested to be  formed by gas-rich major mergers. 
By using stellar and gaseous numerical simulations of galaxy mergers,
we firstly demonstrate that  the central regions of ULIGs are the most promising
formation sites of  super star clusters owing to the rather high gaseous 
pressure of the interstellar medium.
Based on simple analytical arguments,
we secondly discuss the possibility   
that super star clusters in 
an ULIG can be efficiently transferred into the nuclear region 
owing to  dynamical friction and consequently merge with one another  to
form a single compact stellar  nucleus with a seed massive black hole.  
We thus suggest that multiple merging between super  star clusters
formed by nuclear starbursts in the central regions of ULIGs can
result in the  formation of massive black holes.

\end{abstract}


\section{Introduction}
 
Recent high resolution  imaging by $Hubble$ $Space$ $Telescope$ ($HST$)  
has revealed super star clusters (hereafter referred to as SSCs)
with masses of $10^5-10^9$ $M_{\odot}$, very young ages
and large dust extinction   in ultra-luminous infrared galaxies (ULIGs), 
most of which are ongoing mergers (e.g., Shaya et al. 1994
Surace et al. 1998; Surace \& Sanders 1999;
Scoville et al. 2000). 
These SSCs are frequently
observed not only in merging galaxies with and without ultra-luminous infrared activity
(e.g, in the ``Antenna'' by Whitmore et al. 1999) but also in 
starburst galaxies  such as M82 (e.g., de Grijs  et al. 2001),
which implies that the presence of SSCs is  
a generic feature in intensively starbursting or merging galaxies such as   ULIGs
(Lutz 1991; Ashman \& Zepf 1992; Surace \& Sanders 1999).  
These discoveries of SSCs in ULIGs 
have raised the following three questions (e.g., Shaya et al. 1994;
Forbes, Brodie, \& Grillmair 1997; McLaughlin 1999; 
Larsen \& Richtler 2000):
(1) What is the dominant mechanism of SSC formation ?
(2) How can the global dynamics of merging galaxies be associated with
the SSC formation ?
(3) Is the  formation efficiency of SSCs  different
between merging galaxies and normal spirals ? 

The purpose of this Letter is to investigate the formation and dynamical evolution
of SSCs in ULIGs formed by gas-rich major mergers  
and thereby to try and  provide some plausible answers for the above three questions.
In particular, we adopt the  plausible assumption  
that the rather high pressure of the warm interstellar gas 
($P_{\rm g}$ $>$ $10^5$ $k_{\rm B}$: $k_{\rm B}$ is Boltzmann's constant) 
can induce  global collapse of giant molecular clouds 
to form massive  compact  star clusters 
corresponding to SSCs (Jog \& Solomon 1992;  Harris \& Pudritz 1994;
Elmegreen \& Efremov 1997).
Accordingly  we investigate numerically the time evolution of gaseous pressure
in gas-rich mergers with different merger parameters.
Previous theoretical results have already demonstrated
that if a compact stellar system has a very large  number of
stellar-mass black holes left after the normal processes of stellar evolution (in massive stars),
these black holes can evolve into  a single massive black hole
(e.g., Quinlan \& Shapiro 1989; Lee 1995).
Based on simple analytic arguments,
we  suggest that 
SSCs formed in  an ULIG can be transformed   
into  a compact self-gravitating
stellar system (i.e., a considerably large SSC) either  with a few tens  
massive black holes (MBHs) with
masses ranging from $10^4$ to $10^6$  $M_{\odot}$
or with  $10^6$--$10^7$  stellar-mass black holes.   
Thus the present  numerical results combined  with  analytical ones  suggest 
that  the dynamical evolution of SSCs in  
the central regions of ULIGs can be closely associated with the formation
of super-massive black holes.

\section{Model}

Our numerical methods for modeling chemodynamical
evolution of dusty starbursts associated with  galaxy
mergers 
have already been described  by Bekki \& Shioya (1998) 
Furthermore the details of the adopted TREESPH, which
is based on that used in early numerical works (e.g., Heller \& Shlosman 1994), 
are given by Bekki (1995).
We accordingly give these numerical techniques only a brief
review here.
We construct models of galaxy mergers between gas-rich 
disks by using the model of Fall-Efstathiou (1980).
The total mass and the size of a progenitor exponential disk are 
$M_{\rm d}$ and $R_{\rm d}$, respectively. From now on, all the masses and 
lengths are measured in units of $M_{\rm d}$ and  $R_{\rm d}$,
respectively, unless otherwise specified. Velocity and time are 
measured in units of $v$ = $ (GM_{\rm d}/R_{\rm d})^{1/2}$ and
$t_{\rm dyn}$ = $(R_{\rm d}^{3}/GM_{\rm d})^{1/2}$, respectively,
where $G$ is the gravitational constant and assumed to be 1.0
in the present study. If we adopt $M_{\rm d}$ = 6.0 $\times$ $10^{10}$ $
\rm M_{\odot}$ and $R_{\rm d}$ = 17.5 kpc as  fiducial values, then $v$ =
1.21 $\times$ $10^{2}$ km $\rm s^{-1}$   and  $t_{\rm dyn}$ = 1.41 $\times$ $10^{8}$
yr, respectively. The dark-to-disk halo mass ratio and the star-to-gas
mass ratio are  set equal to
4.0 and 9.0, respectively.
Bulge component is not included in the present study.
An isothermal equation of state is used for the gas 
with a temperature of $7.3\times 10^3$ K (corresponding to a sound speed 
of 10 km $\rm s^{-1}$).
The initial metallicity,  $Z_{\ast}$,  for each  gaseous
particle in a given galactic radius  $R$  (kpc) from the center
of a disk is given
according to the observed relation $Z_{\ast} = 0.06 \times {10}^{-0.197 \times (R/3.5)}$
of typical late-type disk galaxies (e.g., Zaritsky, Kennicutt, \& Huchra 1994). 
We present the results mostly  for  merger models with 
nearly prograde-prograde orbital configurations,  parabolic encounters,
and a  pericentric distance of 0.5 $R_{\rm d}$. 
We change the mass ratio ($m_{2}$) of the  two merging disks
and thereby investigate the dependence of SSC formation processes on $m_{2}$
over the range  0.1 (corresponding to minor merging) $\le$ $m_2$ $\le$ 1.0 (major). 
For comparison, we also show the results of an isolated disk model
and a tidally interacting one  (in which two galaxies do not merge with
each other owing to the  large pericentric radius of 2 $R_{\rm d}$
adopted for the   parabolic orbit). Total particle number in each simulation is  23178
for the collisionless components
and 20000 for the  collisional ones.

We adopt the following two different star formation laws  and thereby
convert the gaseous components  into collisionless new stellar particles.
In the first law,  
we  adopt the formation model for  SSCs (globular clusters)  of  Harris \& Pudritz (1994)
in which interstellar gaseous pressure ($P_{g}$) in star forming regions of a galaxy
can drive  pressure confined, 
magnetized self-gravitating  proto-cluster molecular clouds
to collapse to form SSCs if $P_{g}$ is larger than the surface pressure ($P_{\rm s}$)
of the clouds: 
\begin{equation}
P_{\rm g} \ge  P_{\rm s} \sim 2.0\times 10^5 k_{\rm B}. \;
\end{equation}  
In the simulations, 
a gas particle 
becomes $one$  new SSC if the particle's pressure
is larger than $P_{\rm s}$ = $2.0\times 10^5 k_{\rm B}$. 
Although we cannot investigate the detailed physical processes of  SSC formation
in the present $global$ (from $\sim$ 100 pc to 10 kpc scale) simulation,  
the adopted  phenomenological approach of SSC  formation 
enables us to trace at least the most promising
formation sites of SSCs in an admittedly  plausible way.
The other  is the Schmidt law (Schmidt 1959) with the exponent of 1.5
(Kennicutt 1989) and newly formed stars due to the Schmidt law
are referred to as ``normal stars'' (represented by NS from now on) for convenience.
If both conditions are satisfied above, only SSC is assumed to be formed.
The effects of radiative cooling and those of supernovae feedback effects
on SSC formation will be explored in our future papers.
Furthermore, if $P_{\rm s}$ = $2.0\times 10^6  k_{\rm B}$,
the formation efficiency is $\sim$ 6 times smaller.

\begin{figure}
\epsscale{0.5}
\plotone{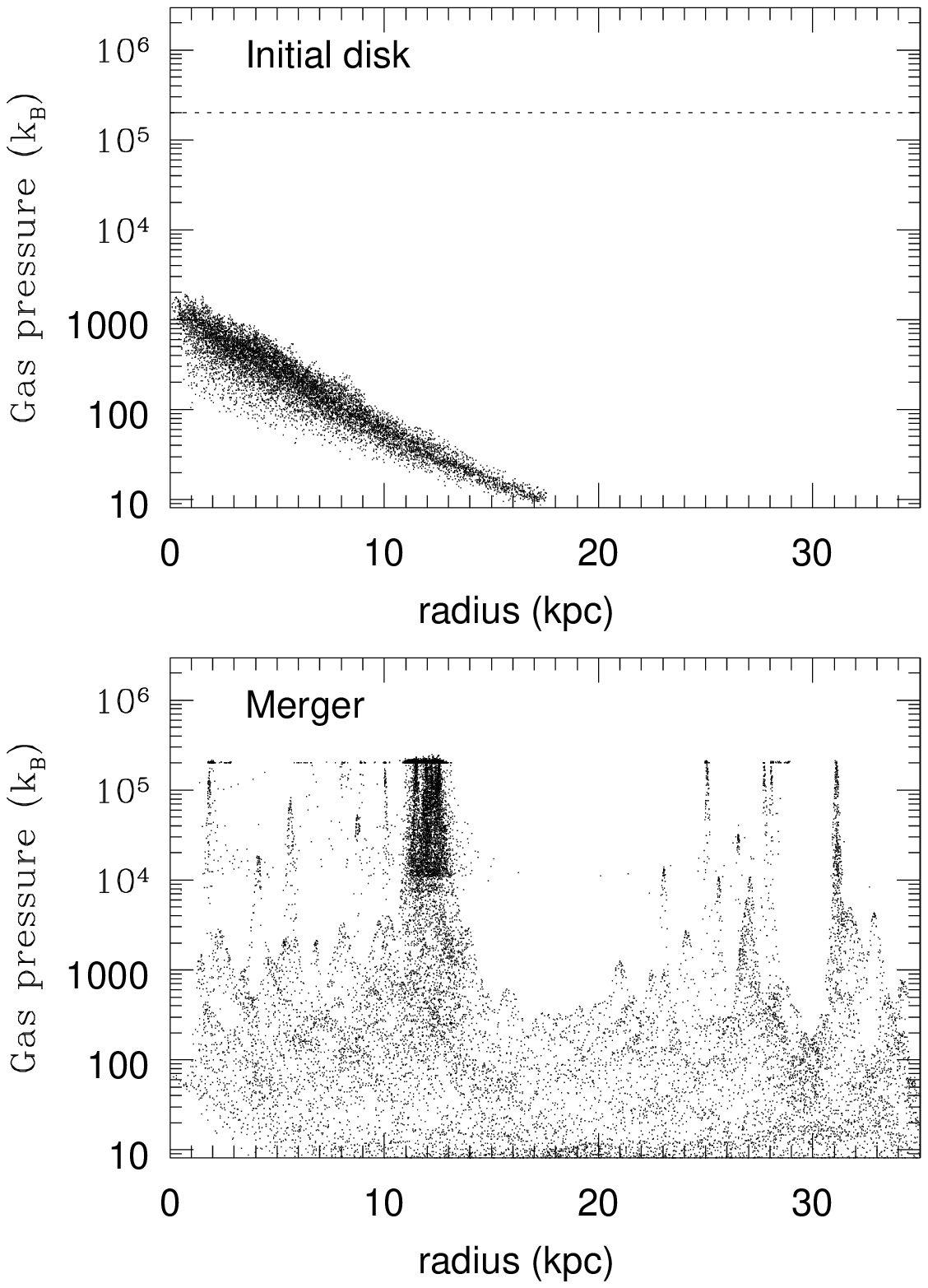}
\caption{
$Upper$:  Distribution of gaseous particles  
on a radius-pressure plane  for the initial disk model.  
The dotted line  shows the threshold pressure over which
SSCs are assumed to be formed. 
$Lower$:  Distribution of  particles that are $initially$ $gaseous$ $particles$  
on a radius-pressure plane  
for the major merger model 
with $m_{2}$ = 1.0 at $T$ = 0.28 Gyr 
(where $T$ represents the time that has elapsed since the two disks begin to merge). 
Here the ``radius'' means the distance
from the center of mass of the merger. 
Not only gaseous particles but also new stellar ones  (SSC and NS)
formed before $T$ = 0.28 Gyr are plotted.
For each new stellar particle, the gaseous pressure  at the epoch when the precursor
gaseous particle is converted into the new stellar one is plotted. 
Accordingly, by comparing the upper panel 
with the lower one, we can clearly observe
how drastically global dynamical evolution
of major galaxy merging has increased the gaseous pressure  until $T$ = 0.28 Gyr. 
The regions around the radius of $\sim$ 12 kpc represent the central starburst cores
in the merger. 
\label{fig-1}}
\end{figure}

\begin{figure}
\epsscale{0.5}
\plotone{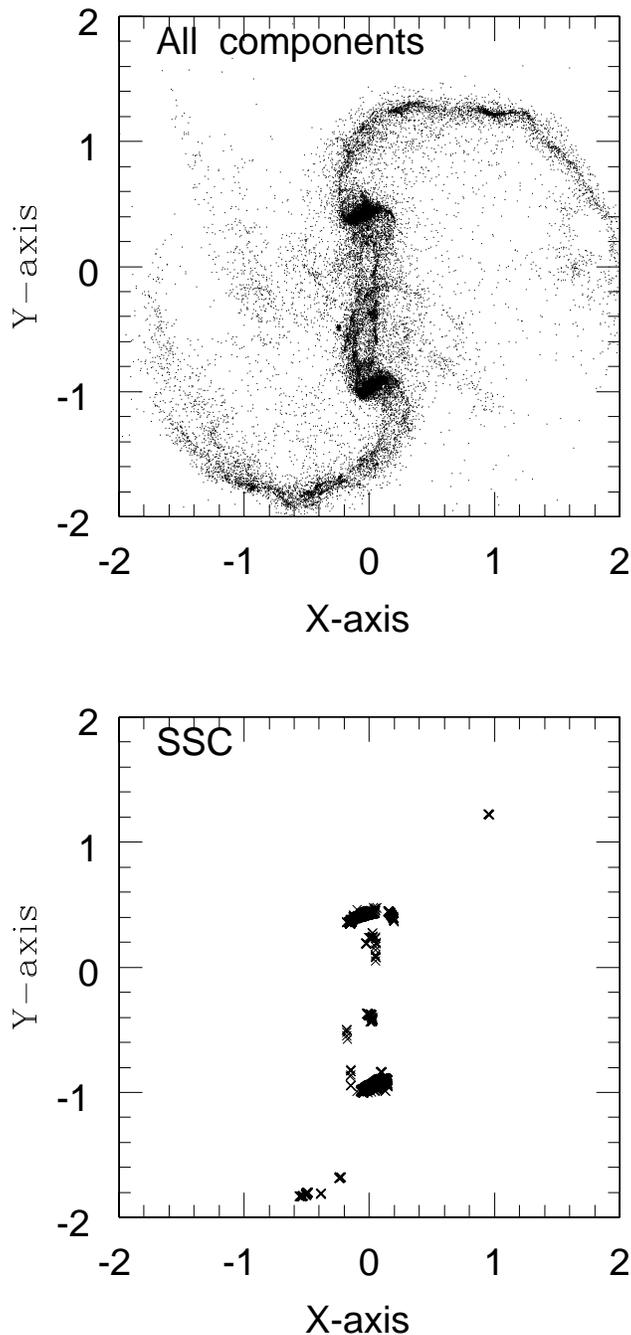}
\caption{
Mass distribution projected onto the $x$-$y$ plane in  the merger model 
with  $m_{2}$ = 1.0 at $T$ = 0.28 Gyr for all components (upper) and
for  SSCs (lower). 
In total, 2667 SSCs have been already formed prior to  this starburst epoch  
(with the star formation rate of $\sim$ 75 $M_{\odot}$ ${\rm yr}^{-1}$).
The scale is given in our units (17.5 kpc), and each of the frames
measures 70 kpc on a side. 
\label{fig-2}}
\end{figure}

\begin{figure}
\epsscale{1.0}
\plotone{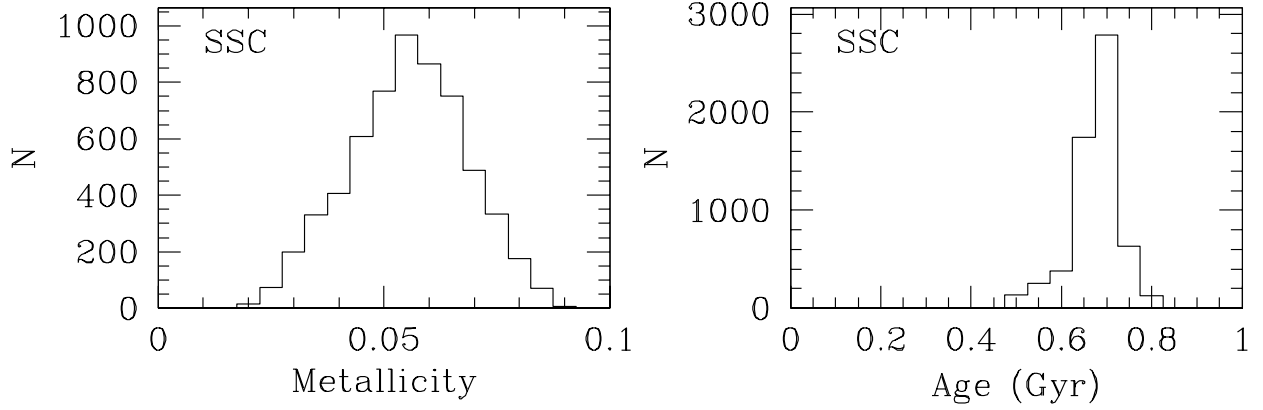}
\figcaption{
Metallicity (left) and age (right) distributions of SSCs in the major
merger model with $m_{2}$ = 1.0 at $T$ = 1.0  Gyr. 
\label{fig-3}}
\end{figure}

\begin{figure}
\epsscale{1.0}
\plotone{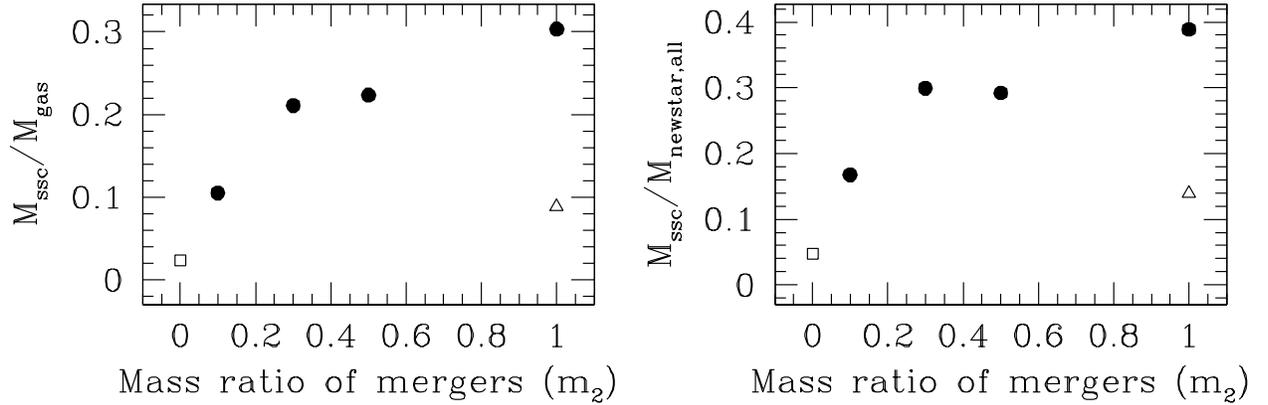}
\figcaption{
The mass ratio of SSCs and initial gas ($M_{\rm ssc}/M_{\rm gas}$ in the left panel)
and the ratio of total mass of SSCs  and that of new stellar components
including both SSCs and NS ($M_{\rm ssc}/M_{\rm newstar,all}$ in the right panel)
for merger models with different $m_{2}$ =0.1, 0.3, 0.5, and 1.0  (filled circles)
at $T$ = 2.0 Gyr.
For comparison, the results for  the isolated disk  and  the interacting  galaxy 
are given by an open square and  an open triangle, respectively.
\label{fig-4}}
\end{figure}

\placefigure{fig-1}
\placefigure{fig-2}
\placefigure{fig-3}
\placefigure{fig-4}

\section{Results}

Figure 1 clearly demonstrates how the  gaseous pressure of the interstellar medium
becomes dramatically higher ($10^4$-$10^5$ $k_{\rm B}$)
in some regions of the major merger model with $m_{2}$ = 1.0 at $T$ = 0.28 Gyr
(starburst phase with the star formation rate of $\sim$ 75 $M_{\odot}$ ${\rm yr}^{-1}$)
in comparison with the initial disks.
Furthermore, gaseous pressure for  some fraction of the gas particles
(in particular, in the  central regions)
exceeds the threshold
value of  $2.0\times 10^5 k_{\rm B}$ for SSC formation,  
essentially because rapid transfer
of gas to the central region of the merger and 
efficient gaseous shock dissipation increase the gaseous density and pressure greatly. 
These results clearly explain  the reason why SSCs can be formed
in ULIGs and imply that the formation efficiency of SSCs are much higher 
in merging galaxies with massive starbursts than in ``normal'' disk galaxies.
Figure 2 shows that SSCs are formed not only in the central region,
where gas fueling is more efficient,  but also
in the outer tidal tails and ``bridges'' among the two cores,
where the tidal force of the merger forms thin and high-density 
gaseous layers  owing to gaseous shock dissipation.  
This result is qualitatively consistent with recent $HST$ observational
results that  SSCs are located 
both in the central regions  (e.g., Arp 220 by Shaya et al. 1994) 
and in  between the two merging cores  (e.g.,  VV114 by Scoville et al. 2000) 
of  ultra-luminous infrared galaxy mergers.

As is shown in Figure 3, the peak metallicity in the metallicity distribution
of SSCs is fairly high (0.05 corresponding to 2.5 solar value and a factor of
2 higher than  metal-rich globular clusters in ellipticals), firstly because
the initial mean metallicity of the merger precursor disk
is assumed to be also high (1.4 solar), and secondly because
chemical enrichment due to starbursts makes the metallicity of SSCs considerably
higher.
We stress here that the derived metallicity distribution
depends strongly on the initial metallicity distributions  of the gaseous disks:
The peak metallicity of SSCs is smaller for a merger with low initial gaseous metallicity. 
Figure 3 also demonstrates that  the age distribution
is rather  narrow (compared with  the distribution in metallicity),
which reflects the fact that  SSC formation is efficient 
only when two disks interact with each other so violently that 
gaseous pressure becomes drastically high ($> 2.0\times 10^5 k_{\rm B}$).

Figure 4 provides the following three predictions on the formation of SSCs.
Firstly, the formation efficiency of SSCs (i.e., the  ratio
of SSC mass to initial gas mass) is more likely to be higher in a merger  
with  larger $m_{2}$. Total SSC number (i.e., formation frequency)
is larger for larger $m_{2}$ and the SSC formation is more widely spread spatially
for larger  $m_{2}$. 
Assuming the same orbital and spin geometry and orientation,
tidal disturbance that takes places in the  galaxy merger
is stronger and the maximum star formation rate is higher for a merger with
larger  $m_{2}$ (Bekki \& Shioya 2000).
Therefore these results imply that a merger with  a more strongly perturbed morphology
and a more massive starburst population can show a higher efficiency of SSC formation.
Secondly, the ratio of the total SSC mass  to that of the new stellar components
(both SSCs and NS) is more likely to be larger 
for a merger with larger $m_{2}$. 
This implies that the luminosity fraction of SSCs, which is observationally estimated
to be less than 20 \% for ULIGs (e.g., Shioya et al. 2001 for the case of Arp 220),   
can be higher
for a merger with larger $m_{2}$.
Thirdly, the higher SSC formation efficiency 
and the larger relative SSC mass ratio 
in major mergers than in tidally interacting galaxies
imply that major mergers, which more dramatically
transform galactic morphologies than tidally  interacting ones (e.g. M82),  
provide the most promising formation
sites of SSCs.

\section{Discussion and conclusion}
One of the most longstanding and remarkable problems
in the formation and evolution of ULIGs  is whether there is an evolutionary
link between the two different types of nuclear activities, starburst and AGN,
(e.g., Sanders \& Mirabel 1996). 
Previous theoretical studies have attempted to understand
either physical mechanisms for the  efficient radial  transfer of gas 
to the nuclear starburst regions in major mergers
(e.g., Barnes \& Hernquist 1992; Mihos \& Hernquist 1996)
or the growth process of the already existing seed massive black holes
(with masses of $\sim$ $10^6$ $M_{\odot}$;
Norman \& Scoville 1988). 
We  discuss here this problem in the context  of
{\it how  a cluster of SSCs in the circumnuclear region of an ULIG 
evolves  dynamically into a seed MBH}.
We suggest that the following  time scale of dynamical friction (Binney \& Tremaine 1987)
is  important for
discussing this problem of SSC evolution:
\begin{equation}
t_{\rm fric}= 3.9\times 10^8(\frac{r_{i}}{500{\rm pc}})^{2}
(\frac{V_{c}}{400 {\rm km s^{-1}}})(\frac{10^7M_{\odot}}{M_{\rm cl}}) {\rm yr}. \;
\end{equation}  
Here we neglect the  ln($\Lambda$) term ($\sim$ 6.6 for a plausible set of parameters),
and $r_{\rm i}$, $V_{\rm c}$, and $M_{\rm cl}$ are the initial radius (from the center
of an ULIG), circular velocity, and mass of the   SSC, respectively.
This time scale argument implies that SSCs formed by dusty starbursts in major mergers
can sink into the central $10-30$  pc region
and consequently merge with one another to  form a single massive nuclear SSC
within a few dynamical time scales of the  ULIG (i.e., $\sim$ $10^8$ yr). 
We here point out  that the fate of SSCs depends strongly on  
whether or not these are tidally disrupted in merging galaxies:
It is also possible that SSCs are disrupted by the central
strong tidal field of ULIGs  before they reach the central few pc in
ULIGs. However, recent observational (e.g., Okuda et al. 1990; Nagata et al. 1995)
and theoretical (e.g., Zwart et al. 2000) results demonstrated that
the central region of the Galaxy can easily harbor compact star clusters 
(corresponding to SSCs).
Although dynamical conditions could be different between the central region
of the Galaxy and those of mergers, 
these results on the Galactic SSCs imply that SSCs in ULIGs can also survive
from tidal disruption owing to their compact nature 
and consequently reach the central few pc via 
dynamical friction.

A high-density compact stellar system has  been demonstrated to form a single MBH 
as a result of the coalescence  of stellar-mass black holes or neutron stars
left after the normal processes of stellar evolution, 
within the dynamical relaxation time scale  of $t_{\rm rel}$ (e.g. Lee 1995). 
Therefore,  if the central region of an ULIG has $\sim$ $10^2$ SSCs with  typical
mass $10^6$ $M_{\odot}$ and if $t_{\rm merge}$ $>$ $t_{\rm rel}$,
a single  very massive nuclear cluster  with $\sim$ $10^2$ seed MBHs 
(each of  mass $\sim 10^5$ $M_{\odot}$) can be formed after
the merging of the SSCs. 
Furthermore, these MBHs in the cluster  may well sink into
the central sub-pc region as a result of  dynamical friction and consequently
form a self-gravitating MBH cluster. Although it could be possible
that multiple merging of these MBHs  through  gravitational  radiation loss
in this MBH cluster can form a single MBH,  the evolution of such multiple
MBH interaction is obviously beyond the scope of this paper
(see more detailed discussions in Valtonen et al. 1994 and Xu \& Ostriker 1994).
On the other hand, if $t_{\rm merge}$ $<$ $t_{\rm rel}$,
then a single massive cluster with huge number of ($\sim$ $10^7$)  stellar-mass black holes
can be formed. In this case also, successive merging  of these black holes
in a cluster can drive the formation of 
a single super massive black hole within the Hubble time
if both the density and the velocity dispersion of the cluster are  rather high 
(Quinlan \& Shapiro 1989). 
Although these discussions are highly speculative,
in each case, 
it seems to be inevitable that
a central supermassive black hole forms  as a natural result
of dynamical evolution of SSCs in the nuclear region of an ULIG.

Recent dynamical studies based on $HST$ photometry and ground-based spectroscopy 
of the central bulge regions of 36 nearby galaxies demonstrated
that the  nuclear MBH (or massive dark object)
to  bulge mass ratio ($f_{\rm BH}$) is  $\sim$ 0.006 (Magorrian et al 1998).
If we adopt the assumption that most giant elliptical galaxies and bulges
are  formed by gas-rich major mergers (e.g., Toomre 1977) and successors
of ULIGs (e.g., Sanders et al. 1988), we can discuss
whether the MBH formation due to merging of SSCs in ULIGs 
is consistent with the observed  $f_{\rm BH}$ value.
For a spheroidal galaxy formed by merging,  
\begin{equation}
f_{\rm BH}=f_{\rm g}f_{\rm sf}f_{\rm ssc}f_{sbh}, \;
\end{equation}  
where $f_{\rm g}$, $f_{\rm sf}$, $f_{\rm ssc}$, and $f_{sbh}$ 
are the gas mass fraction, the mass ratio  of new stars formed
during merging (including SSCs) to gas mass,
that of SSCs
to the total mass of new stars, 
and that of the total mass of stellar-mass black holes  to the total mass
of a SSC. If we adopt a set of plausible values, $f_{\rm g}$ = 0.2 (from the 
typical value observed for  late-type spirals),
$f_{\rm sf}$ = 0.8 (from the present study),
$f_{\rm ssc}$ = 0.2 
(Shioya  et al. 2001), and $f_{sbh}$ = 0.1 (Norman \& Scoville 1988),
then $f_{\rm BH}$ = 0.0032, which is similar to the value observed.
The most important point in the above estimation is that
the origin of $f_{\rm BH}$ can be closely associated with 
{\it the formation efficiency of SSCs in ULIGs} ($f_{\rm ssc}$), 
which has not been pointed out by previous studies.
Finally,  we  suggest that the origin of the physical relationship
between  the masses of MBHs and those of galactic spheroids (e.g., Magorrian et al. 1998) 
can be understood in terms of the formation and evolution of SSCs.

We are  grateful to the anonymous referee for valuable comments,
which contribute to improve the present paper.

\clearpage


\end{document}